\documentclass[aps,prd,twocolumn,nofootinbib,amsmath,amssymb]{revtex4}

\usepackage{graphics}
\usepackage{color}
\begin{document}

\title{Semiclassical partition function for the double-well potential}

\author{D. {\sc Kroff}$^{1,2}$\footnote{krof\mbox{}f@if.ufrj.br},
A. {\sc Bessa}$^{3}$\footnote{abessa@ect.ufrn.br},
C. A. A. {\sc de Carvalho}$^{1}$\footnote{aragao@if.ufrj.br},
E. S. {\sc Fraga}$^{1,4,5}$\footnote{fraga@if.ufrj.br}
and S. E. {\sc Jor\'as}$^{1}$\footnote{joras@if.ufrj.br}}

\affiliation{$^{1}$Instituto de F\'\i sica, Universidade Federal do Rio de Janeiro,
Caixa Postal 68528, 21941-972, Rio de Janeiro, RJ , Brazil\\
$^{2}$Institute de Physique Th\'eorique CEA/DSM/Saclay, 
Orme des Merisiers 91191 Gif-sur-Yvette cedex, France\\
$^{3}$Escola de Ci\^encias e Tecnologia, Universidade Federal do Rio Grande do Norte,
Caixa Postal 1524, 59072-970, Natal, RN , Brazil \\
$^{4}$Institute for Theoretical Physics, J.~W.~Goethe-University, D-60438 Frankfurt am Main, Germany\\
$^{5}$Frankfurt Institute for Advanced Studies, J. W. Goethe University, D-60438 Frankfurt am Main, Germany}

\date{\today}

\begin{abstract}
We compute the partition function and specific heat for a quantum mechanical particle under the influence of a quartic 
double-well potential non-perturbatively, using the semiclassical method. Near the region of bounded motion in the inverted 
potential, the usual quadratic approximation fails due to the existence of multiple classical solutions and caustics. Using the 
tools of catastrophe theory, we identify the relevant classical solutions, showing that at most two have to be considered. 
This corresponds to the first step towards the study of spontaneous symmetry breaking and thermal phase transitions in the 
non-perturbative framework of the boundary effective theory.
\end{abstract}

%%%%%%%%%%%%%%%%%%%%%%%%%%%%%%%%%%%%%%%%

\maketitle

%%%%%%%%%%%%%%%%%%%%%%%%%%%%%%%%%%%%%%%%
\section{Introduction and motivation}

In the analytic description of phase transitions in particle physics and nuclear theory, one usually relies on the effective model 
approach, given the complexity of the fundamental theories involved. 
If we consider strong interactions, the phase diagram related to chiral symmetry restoration and deconfinement is a particularly interesting example, since they are within experimental reach and currently being investigated by different experiments at RHIC-BNL and LHC-CERN \cite{Ullrich:2013qwa}.
Usually, one generally adopts low-energy effective models such as the linear sigma model \cite{GellMann:1960np,lee-book} and the 
Nambu--Jona-Lasinio model \cite{Klevansky:1992qe}, which can be combined with different versions of the Polyakov loop 
model \cite{Pisarski:2000eq}. 
The standard approach, then, corresponds to the computation of a thermal effective potential from 
which one can extract information on the different phases and all thermodynamic quantities, so that one can build a phase diagram.

In most cases, the computation is performed in the mean-field approximation with one-loop thermal corrections assuming 
homogeneous and static background fields \cite{FTFT-books}. Frequently, vacuum loop contributions are ignored, even in a theory 
with spontaneous symmetry breaking, where the presence of a condensate always modifies the masses, which then become medium-dependent 
quantities, affecting significantly the phase structure \cite{Mocsy:2004ab,Palhares:2008yq,thesis,Fraga:2009pi,Boomsma:2009eh,Mizher:2010zb,Skokov:2010sf,Palhares:2010be,Andersen:2011pr,Mintz:2012mz}. 
So, the highly non-linear behavior of the effective potential for large fields is completely missed, as well as non-perturbative effects 
(with the exception of the treatment within the functional renormalization group \cite{FRG}). Those aspects can, in principle, dramatically 
modify the phase structure provided by a given effective model.

The boundary effective theory formalism \cite{Bessa:2010tp,Bessa:2010tj} furnishes a non-perturbative method to 
calculate the partition function of quantum systems in thermal equilibrium in which configurations that are not strictly periodic play the main role. In such approach, one 
can compute the thermal one-loop effective potential for a system of massless scalar fields with quartic interaction \cite{Bessa:2011tn}. 
The calculation relies on the solution of the classical equation of motion for the field, and Gaussian fluctuations around it. The 
result is non-perturbative and differs from the standard one-loop effective potential \cite{Dolan:1973qd} for field values larger 
than $T/\sqrt{\lambda}$, $T$ being the temperature and $\lambda$ the coupling  \cite{Bessa:2011tn}.

The natural extension would be the calculation of the effective potential in the case with spontaneous symmetry breaking.
That would allow for the description of phase transitions in effective models incorporating non-linear and 
non-perturbative effects, as well as controlling the infrared divergences of thermal field theory in a well-defined and relatively simple 
way \cite{Bessa:2010tp,Bessa:2010tj,Bessa:2007vq}.  However, to develop the method to be applied in this case, it is necessary to deal with multiple classical solutions, since more than one solution may satisfy the boundary conditions in euclidean time. 

As a first step towards the study of spontaneous symmetry breaking and thermal phase transitions using the boundary effective 
theory, in this paper we focus on the simpler case of computing the semiclassical partition function for a quartic 
double-well potential in quantum statistical mechanics. Although apparently trivial and straightforward, the inverted potential in 
this case has a region of bounded motion. Therefore, one also has to deal with multiple solutions and their coalescence as the 
temperature changes. The usual quadratic approximation may yield good results when such solutions are far away from each other 
in functional space, but, as we shall see later on, this is not so in the opposite scenario. Among the numerous solutions, we use the 
framework of catastrophe theory to identify the only two relevant ones, following Refs. \cite{deCarvalho:1998ff,deCarvalho:2001vk}. 
We then compute the partition function and specific heat, obtaining the correct limits at both high and low temperatures, and a regular behavior where the usual quadratic approximation diverges.

The paper is organized as follows. In Section II we review general characteristics of the semiclassical path-integral representation 
of the partition function and discuss the case of multiple solutions in the double well. In Section III we use the tools from catastrophe 
theory to deal with the coalescence of solutions and identification of relevant minima. In Section IV we present results for the 
partition function and the specific heat, discussing their controlled behavior and the domain of validity of our approximation. 
Section V contains our summary. Elements and some technical details of catastrophe theory are presented in appendices. 

%%%%%%%%%%%%%%%%%%%%%%%%%%%%%%%%%%%%%%%%

\section{Semiclassical path-integral representation of the partition function}
\subsection{General Features}
In statistical mechanics, the partition function for a system in contact with a thermal reservoir at temperature 
$T$ is given by the sum of a probabilistic weight, the diagonal elements of the density matrix, over a stochastic 
variable that labels the state of the system. This object is of fundamental importance, as it encodes all the 
thermodynamic information.

For a one-dimensional quantum-mechanical system consisting of a single particle, the stochastic variable can be chosen 
as the  Schr\"odinger-picture position operator eigenvalue. Therefore, if the dynamics is dictated by the Hamiltonian 
operator $\hat{H}$, the partition function is written as ($1/\beta \equiv k_BT$):
\begin{equation}
\label{partfunc}
Z = \int_{-\infty}^{\infty} dx_0 \  \langle x_0| \exp(-\beta \hat{H}) |x_0\rangle \ .
\end{equation}

The matrix element in the previous equation can be understood as the analytic continuation of the transition amplitude 
$\langle x_0| \exp[-i(\hat{H}/\hbar)(t-t')]|x_0\rangle$ to imaginary time, allowing for a formal expression for the 
diagonal elements  of the density matrix in terms of path integrals \cite{Feynman}. If we restrict our analysis to systems 
subject to velocity-independent potentials, the desired expression has the well-known form:
\begin{equation}
\label{euclidpint}
\langle x_0|\exp(-\beta\hat{H})|x_0\rangle = \!\!\!\!\!\!\!\!\!\!\!\!\!\!\!\!\!\!\!\!\!\!\!\!\!\!\!\!\int\limits_{\,\,\,\,\,\,\,\,\,\,\,\,\,\,\,\,\,\,\,\,\,\,\,\,\,\,\,\,\,\,\,\,\, x(0) = x(\beta\hbar)=x_0} 
\!\!\!\!\!\!\!\!\!\!\!\!\!\!\!\!\!\!\!\!\!\!\!\!\!\!\![{\cal D} x(\tau)] \ e^{-S_E/\hbar} \ ,
\end{equation}
\noindent
where
\begin{equation}
\label{euclidaction}
S_E[x] = \int_0^{\beta} d\tau\left[ \frac{m}{2} \left(\frac{dx}{d\tau}\right)^2 + V(x)\right] \ .
\end{equation}

\noindent
In other words, the diagonal elements of the density matrix are obtained integrating the exponential of the Eucliden action 
$S_E$ over the paths $x(\tau)$ in imaginary time satisfying the conditions $x(0) = x(\beta\hbar) = x_0$.

For convenience we define the dimensionless quantities $q\equiv x/x_N$, $\theta \equiv \omega_N\tau$, 
$\Theta \equiv \beta\hbar\omega_N$, $U(q) \equiv V(x_Nq)/m\omega_N^2x_N^2$ and 
$g \equiv \hbar/m\omega_Nx_N^ 2$ where $\omega_N^{-1}$ and $x_N$ are the natural time and length scales of the 
problem under consideration, respectively. In terms of these, the partition function can be written as follows:

\begin{equation}
\label{apartfunc}
Z(\Theta) = \int_{-\infty}^{\infty} dq_0 \!\!\!\!\!\!\!\!\!\!\!\!\!\!\!\!\!\!\!\!\!\!\!\!\!\!\!\int\limits_{\,\,\,\,\,\,\,\,\,\,\,\,\,\,\,\,\,\,\,\,\,\,\,\,\,\,\,\,\,\,\,\,\,q(0) = q(\Theta) = q_0} \!\!\!\!\!\!\!\!\!\!\!\!\!\!\!\!\!\!\!\!\!\!\!\!\![{\cal D} q(\theta)] \ e^{-I/g} \ 
,
\end{equation}
\noindent
where
\begin{equation}
\label{aeuclidaction}
I[q] = \int_0^{\Theta} d\theta\left[ \frac{1}{2} \left(\frac{dq}{d\theta}\right)^2 + U(q)\right] \ .
\end{equation}

In general, it is not possible to solve exactly the path integral above, but we can still resort to approximation 
procedures in order to evaluate it. A very natural approach is the JWKB \cite{Schulman} asymptotic expansion in $\hbar$ (or $g$) --- also known as semiclassical approximation --- that we briefly discuss below.

The trajectories that extremize the euclidean action $I[q]$ are those satisfying the Euler-Lagrange equation 
($U' \equiv dU/dq$),
\begin{equation}
\label{classicalsol}
\frac{d^2q_c}{d\theta^2} - U'(q_c) = 0 \ ,
\end{equation}

\noindent
subject to the boundary conditions $q(0) = q(\Theta) = q_0$. In other words, these are the classical solutions 
describing the motion of a particle under the influence of the inverted potential $-U(q)$. 

Due to its euclidean nature, the path integral in equation \eqref{euclidpint} is dominated by the functions in the 
vicinity of those that {\it minimize} $I[q]$. So, one has to determine among the solutions of \eqref{classicalsol} 
those representing minima, which we denote by $\bar{q}_c^{~i}$. Expanding the action around the minima, we have 
$I[\bar{q}_c^{~i} + \eta] = I[\bar{q}_c^{~i}] + I_2[\bar{q}_c^{~i},\eta] + \delta I[\bar{q}_c^{~i}, \eta]$, where
\begin{equation}
\label{quadfluct}
I_2[\bar{q}_c^{~i},\eta] = \frac{1}{2}\int_0^{\Theta} d\theta \ \eta(\theta)\left[-\frac{d^2}{d\theta^2} + 
U''(\bar{q}_c^{~i})\right]\eta(\theta) \ ,
\end{equation}

\begin{equation}
\label{higherfluct}
\delta I[\bar{q}_c^{~i}, \eta] = \sum_{k = 3}^{\infty}\frac{1}{k!}\int_0^{\Theta} d\theta \ U^{(k)}
(\bar{q}_c^{~i})\eta^k(\theta) \ .
\end{equation}

\noindent
Keeping only terms up to quadratic order in the fluctuations, one obtains the so-called {\it standard}  semiclassical 
approximation for the partition function:
\begin{equation}
\label{semiclassZ}
Z \approx \int_{-\infty}^{\infty} dq_0 \sum_i \exp(-I[\bar{q}_c^{~i}]/g) \Delta ^{-1/2} \ ,
\end{equation}
\noindent
where $\Delta$ is the determinant of the quadratic fluctuation operator
\begin{equation}
\label{determinant}
\Delta = \det \hat{F}[\bar{q}_c^ {~i}] = \det\left[-\frac{d^2}{d\theta^2} + U''(\bar{q}_c^{~i})\right] \ .
\end{equation}

\begin{figure}[ht]
\begin{center}
\resizebox*{!}{5.0cm}{\includegraphics{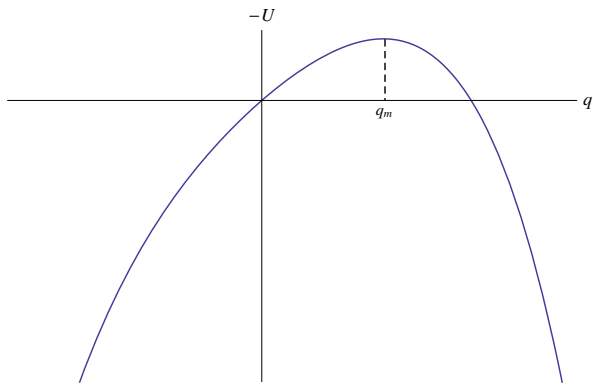}}
\end{center}
\vspace{-5mm}
\caption{\label{figsingle} Single-well inverted potential.}
\end{figure}

As an example, let us consider a single-well potential $U(q)$, whose global minimum is located at the 
point $q_m$, as depicted in Fig. \ref{figsingle}. Following the method described above, one has to obtain the solutions 
describing the classical motion of the particle under the influence of the inverted potential that leaves the point $q_0$ at 
$\theta = 0$ and returns after a time interval $\Theta$.

As the potential $-U(q)$ is unbounded from below, if the particle departs from a point such that $q_0 < q_m$ 
($q_0 > q_m$), it will only return to the initial position if its initial velocity points to the right (left), 
otherwise the particle will move directly towards  $-\infty$ ($+\infty$). However, the initial velocity can not be arbitrarily large, for if the particle energy is greater than the height of the potential barrier, it will not return to its initial position either, as it will move directly towards $+\infty$ ($-\infty$). Thus, the maximum possible value for the particle energy is exactly the barrier height. 

For a fixed value of $q_0$, the time the particle spends going from the initial position up to the turning point 
$q_t$ is a function of $q_t$ only, given by the following expression:
\begin{equation}
\label{ToF}
\frac{\Theta}{2} = \textrm{sign}(q_t - q_0)\int_{q_0}^{q_t}\frac{dq}{\sqrt{2[U(q)-U(q_t)]}} \ .
\end{equation}
Clearly, the previous expression vanishes when $q_t = q_0$. But, as the turning point moves further up the barrier, 
the time of flight increases continuously, diverging when $q_t$ is exactly at the top. Therefore, for any value of $\Theta$, it is 
possible to determine {\it the one solution} satisfying $q(0) = q(\Theta) = q_0$ by choosing the appropriate 
turning point. It is, then, a straightforward task to implement the semiclassical method, as was demonstrated in 
detail in Ref. \cite{deCarvalho:1998mv}. In fact, even the $D$-dimensional case can be treated for central 
potentials \cite{deCarvalho:1999fi}.

%
%%%%%%%%%%%%%%%%%%%%%%%%%%%%%%%%%%%%%%%%

\subsection{Double-well potential: multiple solutions}
\begin{figure}[ht]
\begin{center}
\resizebox*{!}{5.0cm}{\includegraphics{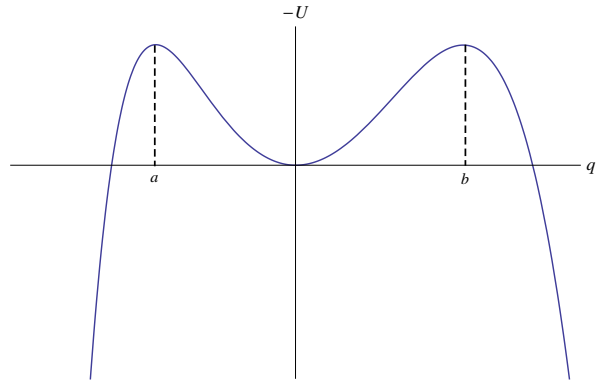}}
\end{center}
\vspace{-5mm}
\caption{\label{figdouble} Double-well inverted potential.}
\end{figure}

The problem becomes more intricate in the case of double-well potentials. Suppose now that $U(q)$ represents a 
double-well potential with degenerate minima located at $q = a$ and $q = b$, with $a < b$, e.g. like the one sketched 
in Fig. \ref{figdouble}. As we shall see, it is now necessary to deal with multiple classical solutions\footnote{The quartic double-well 
potential at finite temperature has been investigated previously using semiclassical, variational and perturbative 
methods, e.g. in Refs. \cite{Harrington:1978tx,Dolan:1979qx,firenze,Bachmann:1998pz}.}.

If $q_0 < a$ or $q_0 > b$, the particle lies in a region of unbounded motion under the potential $-U$, resembling the single-well case.\footnote{It is essential for this piece of the argument that the maxima of the inverted potential are degenerate. The method discussed in the present work can be generalized and applied to the case of non degenerate minima.} It is trivial to extend the 
arguments given in the previous section and conclude that, as before, each pair $(q_0,\Theta)$ defines a unique solution to equation \eqref{classicalsol}.

Let us now analyze what happens when $ a < q_0 < b$, i.e. when the particle starts in the well of $-U(q)$. Once again, 
its energy has to be smaller than the barrier height, otherwise the particle will leave the well towards 
$\pm\infty$ without ever returning to its initial position. In other words, there is a maximum allowed speed for such particles and all the solutions departing 
from a point in the well must always remain therein. In this region of bounded motion, we see a much richer structure, with the possibility of multiple 
classical solutions for a given $q_0$, depending on $\Theta$. 

If the temperature is high enough, the available time of flight is still very restrictive. Accordingly, since the speed is limited, the particle 
will be able to move only towards the nearest peak ($q_t$ and $q_0$ will have the same signal --- see Fig. \ref{figrib}, lower panel), but it
will not  be able to reach points too far from its initial position. Thus, in this limit, we still have a single solution for 
every $q_0$. Lowering the temperature (increasing $\Theta$), the particle will be able to go further away and eventually it will be able to reach 
also the opposite side of the potential well and return to its initial position. From then on (i.e, for lower temperatures), a fixed $q_0$ will no longer 
define a unique classical solution \cite{deCarvalho:1998ff}.

In order to apply the semiclassical method with multiple solutions, one has to be able to identify and keep only those representing minima of the euclidean action in 
functional space, discarding maxima and saddle points, which both correspond to unstable solutions with at least one negative eigenvalue\footnote{The solution 
$q_c(\theta)$ is a minimum when all the eigenvalues are positive, a maximum when they are all negative and a saddle-point otherwise.} of  the quadratic fluctuation 
operator $\hat{F}[q_c]$, defined in equation \eqref{determinant}. 

In the next section, we restrict ourselves to a quartic double-well potential and, using the language of 
catastrophe theory, we not only identify how the number of solutions changes as we vary the parameters $(q_0,\Theta)$, 
but also find a straightforward criterium to determine which classical trajectories must be taken into account.

%%%%%%%%%%%%%%%%%%%%%%%%%%%%%%%%%%%%%%%
%
\section{Coalescence of solutions} %figs and text for dummies!
\subsection{Caustics and catastrophes for the quartic double-well potential}
From now on, we consider the specific case of a quartic double-well potential, $V(x) = -m\omega^2x^2/2 + \lambda 
x^4/4$. Writing it in terms of $q \equiv x/x_N$, $x_N \equiv \sqrt{m\omega^2/\lambda}$:
\begin{equation}
\label{quarticdwell}
U(q) = \frac{\lambda}{m^2\omega^4}V(x_Nq) = -\frac{1}{2}q^2 + \frac{1}{4}q^4 \ .
\end{equation} 

As discussed previously, if the temperature is sufficiently high, there is only one closed path, with a single turning 
point, for every $q_0$. Lowering the temperature, we go from a single-solution to a three-solution regime. Lowering it 
even further, we reach a five-solution regime and so on \cite{deCarvalho:1998ff}.

\begin{figure}[ht]
\begin{center}
\resizebox*{!}{10.0cm}{\includegraphics{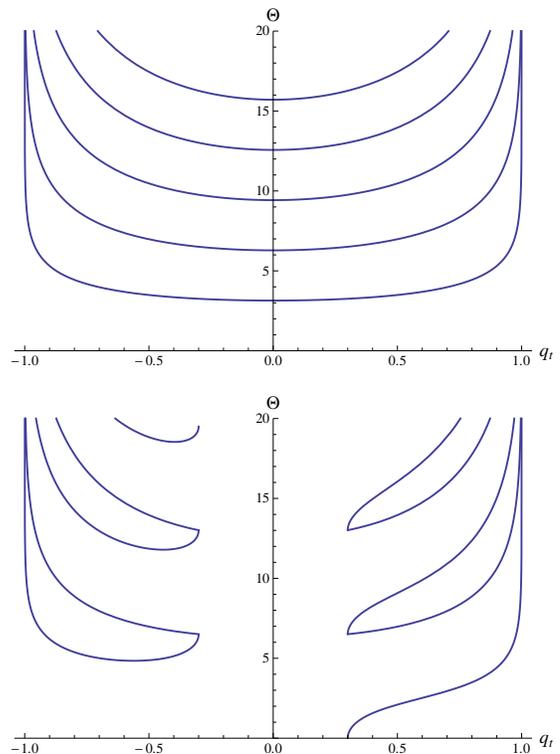}}
\end{center}
\vspace{-5mm}
\caption{\label{figrib}Plots of the time of flight $\Theta$ vs the turning point $q_t$ for $q_0 = 0$ (upper) and $q_0 = 0.3$ (lower). In the upper panel, the trivial solution 
$q_0=q_t=0=q(t) \, \forall t$, although valid for all $\Theta$, is not shown.}
\end{figure}

Fig. \ref{figrib} is a clear illustration of the feature of solution bifurcation. It shows the plots of the time of flight $\Theta$ {\it vs.} the first turning point $q_t$ of 
the classical path for two different values of $q_0$. One can read directly from the plots the values of $q_t$ that allow the particle to return to $q_0$ in a time 
interval $\Theta$. As fixing the initial position and the first turning point defines univocally the classical trajectory, the plot shows the number of classical 
solutions related with each value of the time of flight.

\begin{figure}[ht]
\begin{center}
\resizebox*{!}{5.0cm}{\includegraphics{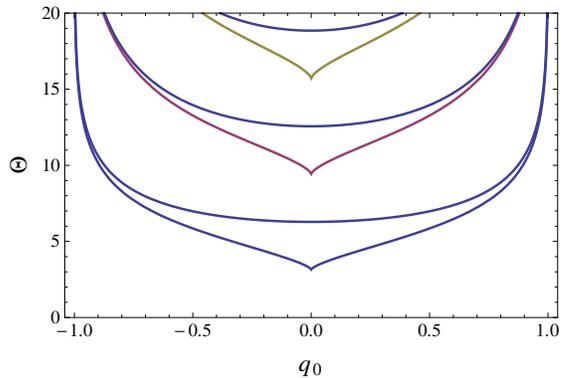}}
\end{center}
\vspace{-5mm}
\caption{\label{figcaustics}The plane $(q_0,\Theta)$ is divided into regions with different number of classical solutions. The
frontiers between those regions, named caustics, are shown above. The smooth curves indicate the coalescence of strictly periodic solutions, i.e, 
those that begin and end at the same position and at the same velocity.}
\end{figure}

Thus, the plane $(q_0,\Theta)$ is divided into several regions with different numbers of solutions, as shown in Fig. \ref{figcaustics}. Moving from a certain region to a 
neighbouring one, two solutions are either created or annihilated. Exactly at the frontier between those regions, two classical trajectories coalesce. The curves defining 
the frontiers between two such regions are named {\it caustics}, for they are analogous to the optical phenomenon.  In our case, the classical solutions play the role of 
the light rays and the action replaces the optical distance \cite{deCarvalho:1998ff}.

\begin{figure}[ht]
\begin{center}
\begin{tabular}{cc}
\multicolumn{2}{c}{  
\resizebox*{9.0cm}{5.0cm}{\includegraphics{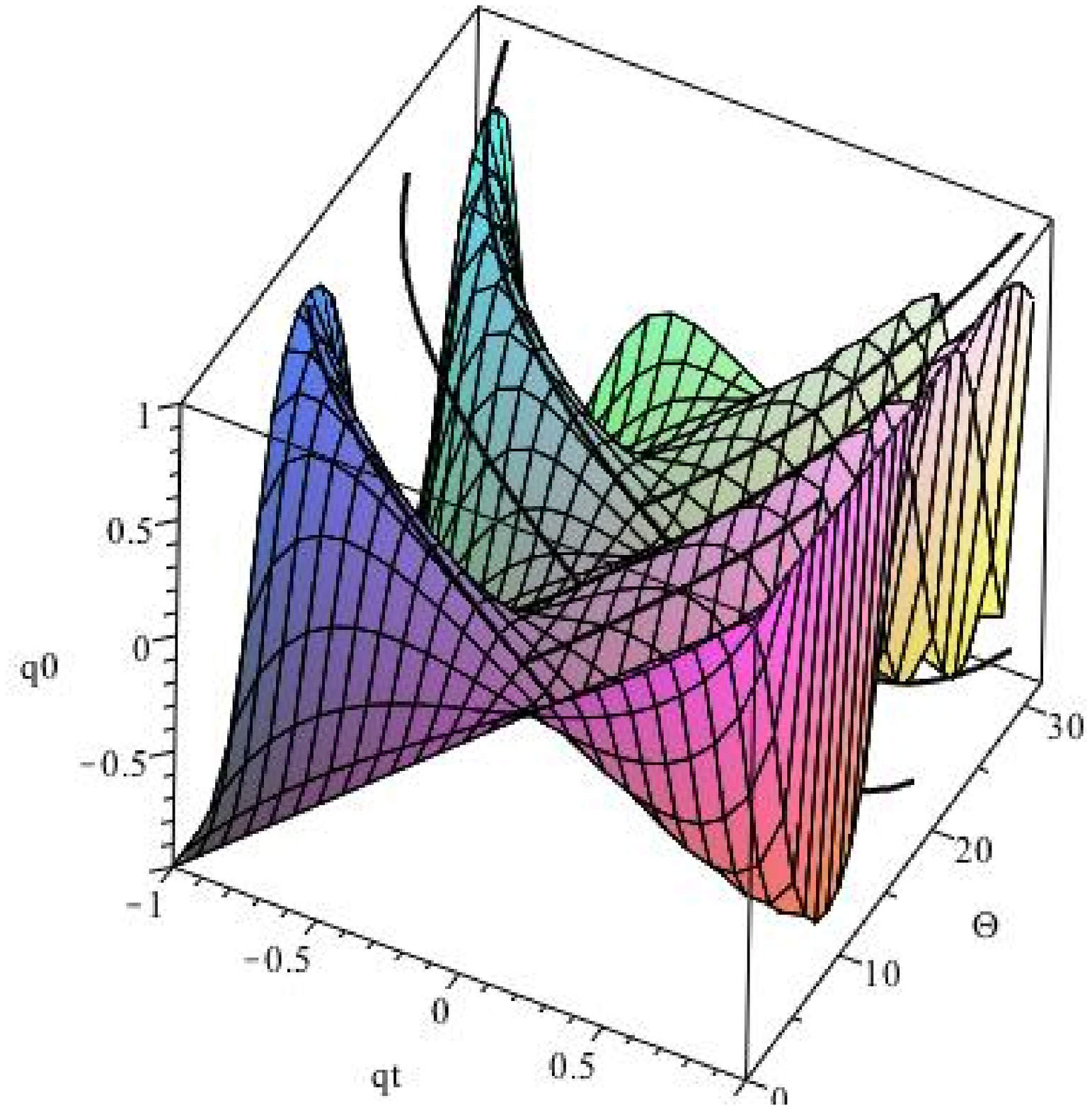}} }\\ 
\resizebox*{!}{4.0cm}{\includegraphics{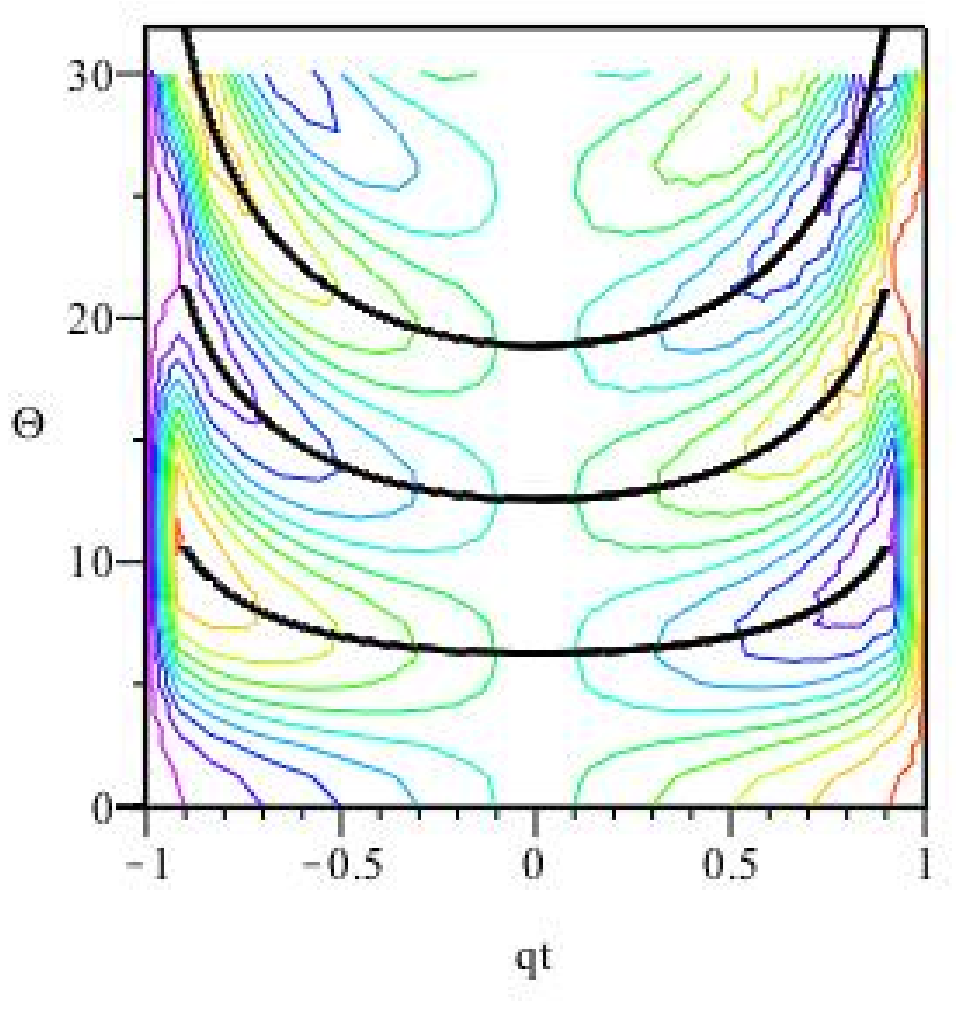}}  &
\resizebox*{!}{4.0cm}{\includegraphics{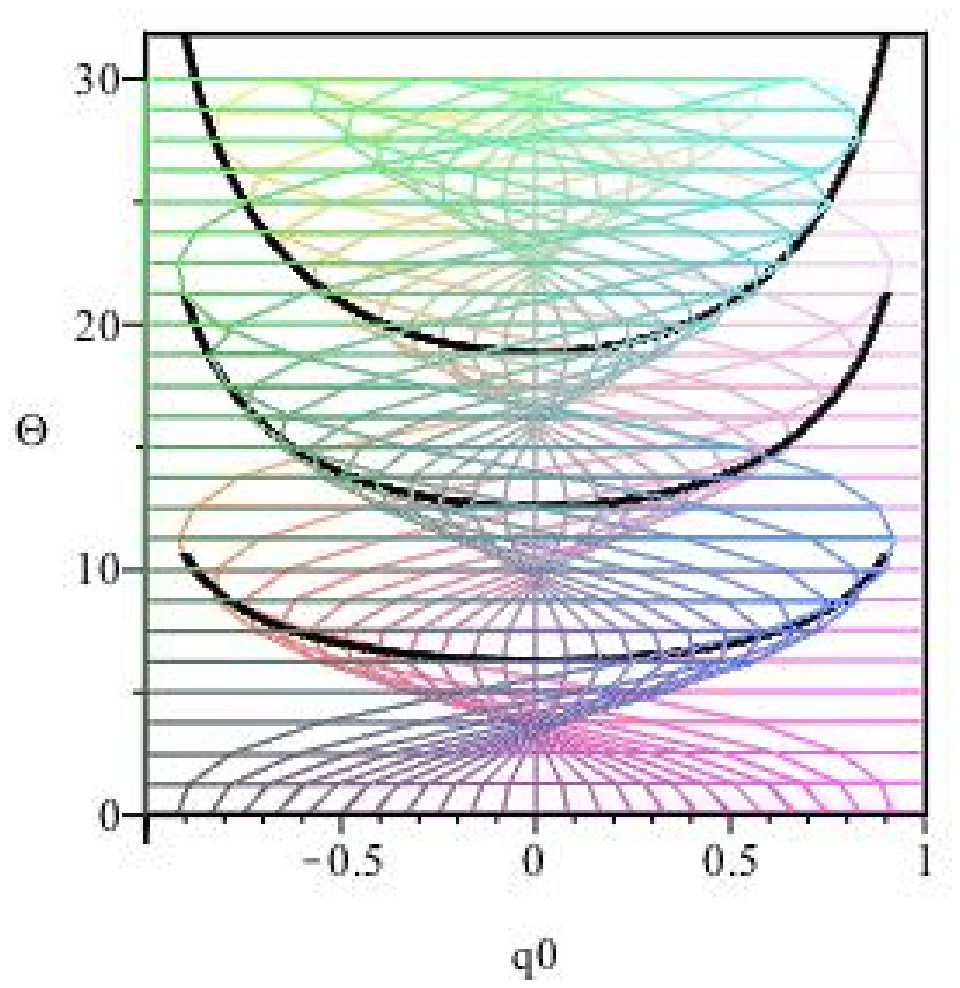}} \\
\end{tabular}
\end{center}
\vspace{-5mm}
\caption{\label{3d} Three dimensional plot of $q_t \times q_0 \times \Theta$ (upper panel) and its projections onto the planes $q_t\times\Theta$ 
(bottom left) and $q_0 \times \Theta$ (bottom right).}
\end{figure}

The information depicted in Figures \ref{figrib} and \ref{figcaustics} is combined into a single 3D plot in Figure \ref{3d}.

To apply the semiclassical method to compute the partition function for the double-well potential we must determine 
which of the solutions are actual minima of the action and track them down. To do so, we use the 
framework of catastrophe theory \cite{catastrophe}, which classifies and studies how the extrema of certain 
functions coalesce and emerge. A brief summary of a few ingredients of catastrophe theory is presented in 
Appendix~\ref{appcatastrophe}.

\subsection{Finding the minima}
Following Appendix~\ref{appcatastrophe}, new real solutions of \eqref{classicalsol}, i.e. new extrema of the euclidean 
action, emerge whenever the fluctuation determinant $\Delta$ vanishes. This provides a criterium to determine the 
location of the caustics.

It is clear that, for the problem under consideration, there are only two variables controlling the pattern of action 
extrema, which we choose to be $q_0$ and $\Theta$.\footnote{The third available quantity, $q_t$, can be written in terms of the chosen ones.} Therefore, we 
are dealing with catastrophes whose codimension is not greater than two. The only two catastrophes satisfying this condition are the fold and the cusp, both 
having only one essential variable or coordinate \cite{catastrophe}. In other words, we know that, in our case, only one eigenvalue of the fluctuation operator 
vanishes when a caustic is crossed. 

\begin{figure}[ht]
\begin{center}
\resizebox*{!}{7cm}{\includegraphics{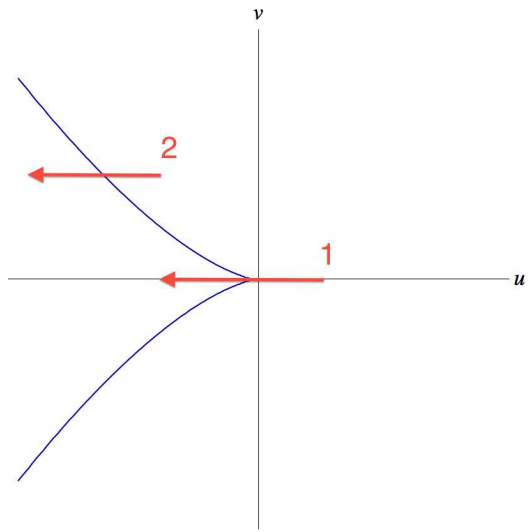}}
\end{center}
\vspace{-5mm}
\caption{\label{figcuspid}Cuspid $27v^2+4u^3=0$ in the control parameter space. See text and Figs. \ref{figaction1} and \ref{figaction2} for explanation of the arrows.}
\end{figure}

Therefore, we can focus on one direction of the functional space: the one defined by the eigenfunction whose lowest eigenvalue
vanishes. If we project the action onto that direction and perform a change of variables (see Appendix \ref{appnormal}), 
we will reach the so-called {\it normal form} (see Table \ref{tablenormalform}):
\begin{equation}
I_N(z) = \frac{1}{4}z^4 + \frac{u}{2}z^2 + vz + s \ ,
\label{normalform}
\end{equation}
\noindent
where $z$ is the coordinate associated to the aforementioned direction in functional space and $u$ and $v$ are the control parameters. The bifurcation set is then given by
\begin{equation}
\frac{dI_N}{dz} = \frac{d^ 2I_N}{dz^2} = 0 \ \Rightarrow \ 27v^2+4u^3 = 0 \ .
\label{bifeq}
\end{equation}
The previous equation defines a cusp in the control parameter space $(u,v)$, dividing it in two parts --- see Fig. \ref{figcuspid}. To the right 
of the curve the action has one minimum, to its left there are one maximum and two minima.

\begin{figure}[ht]
\begin{center}
\resizebox*{!}{1.8cm}{\includegraphics{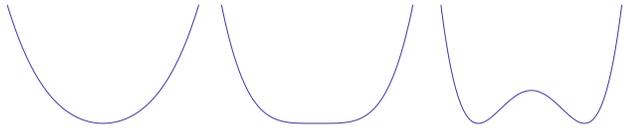}}
\end{center}
\vspace{-5mm}
\caption{\label{figaction1}Behavior of the action in functional space when the cusp is crossed at the vertex. See arrow $1$ in Fig. \ref{figcuspid}.}
\end{figure}

\begin{figure}[ht]
\begin{center}
\resizebox*{!}{1.9cm}{\includegraphics{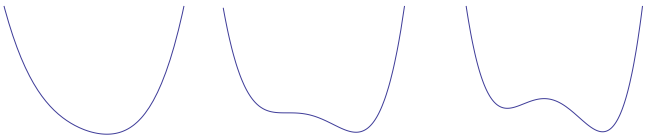}}
\end{center}
\vspace{-5mm}
\caption{\label{figaction2}Behavior of the action in functional space when the cusp is crossed at a point that 
is not the vertex.  See arrow $2$ in Fig. \ref{figcuspid}.}
\end{figure}

If the cusp is crossed at its vertex, as in arrow $1$ of Fig. \ref{figcuspid}, the original minimum becomes a maximum and two symmetric minima appear --- see Fig. 
\ref{figaction1}. On the other hand, if the crossing happens at any other point, as in arrow 2 in Fig. \ref{figcuspid}, the original minimum remains and two new solutions 
appear --- a maximum and a new (local) minimum --- see Fig. \ref{figaction2}. In the former case, one solution splits into three; in the later, two new solutions emerge 
(out of the coalescence of their complex counterparts --- see next paragraph) while the previously existing minimum is unaffected. This picture agrees with our 
previous statement that the number of solutions of \eqref{classicalsol} increases by two.

It is useful to think of exactly the same merging of extrema that happens in the algebraic equation (\ref{normalform}). Being a fourth-order polynomial with real 
coefficients, there are always three extrema which may be either real or imaginary, depending on the values taken by the control parameters $\{u,v\}$. Then, one usually 
speaks of the coalescence of {\it complex} solutions (which always come in pairs and are conjugate to each other) and their subsequent separation along the real axis, 
as opposed to their plain {\it creation} out of nothing.

As one lowers the temperature, at the next catastrophe the classical trajectory with highest action is the one that gives rise to two new solutions. 
So, the solutions emerging after the second caustic are still maxima along the direction (in functional space) of the first catastrophe. On all the forthcoming 
catastrophes, the same happens: new solutions originate from the one with the highest action.
Thus, in the multiple-solution regime, the only solutions that are actual minima of the action are those that are 
minima along the direction of the first catastrophe. So, to apply the semiclassical method, we just have to be 
concerned about at most two classical solutions. 

Moreover, as the first catastrophe happens before the emergence of strictly periodic solutions (in fact, these solutions appear only in the second catastrophe), we can 
guarantee that the solutions we have to keep have a single turning point.
Hence, there is a criterium that allows to determine which solutions of \eqref{classicalsol} we must use when applying the semiclassical method: {\it at the end we need only the single-turning-point trajectories}.

%%%%%%%%%%%%%%%%%%%%%%%%%%%%%%%%%%%%%%%
%
\section{Partition function and specific heat for the double-well potential}
The solutions of the classical equation of motion for the potential given in equation \eqref{quarticdwell} can be 
expressed in terms of Jacobi Elliptic Functions \cite{Byrd,Gradshteyn}. In particular, the solutions we are interested 
in, the ones with a single turning point, can be written as
\begin{equation}
q(\theta) = q_t \textrm{cd}\left[\sqrt{1-q^2_t/2}(\theta-\Theta/2),k\right]
\label{jacobicd}
\end{equation}

\noindent
where we define $k \equiv q_t/\sqrt{2-q^2_t}$.

\begin{figure}[ht]
\begin{center}
\resizebox*{!}{10.0cm}{\includegraphics{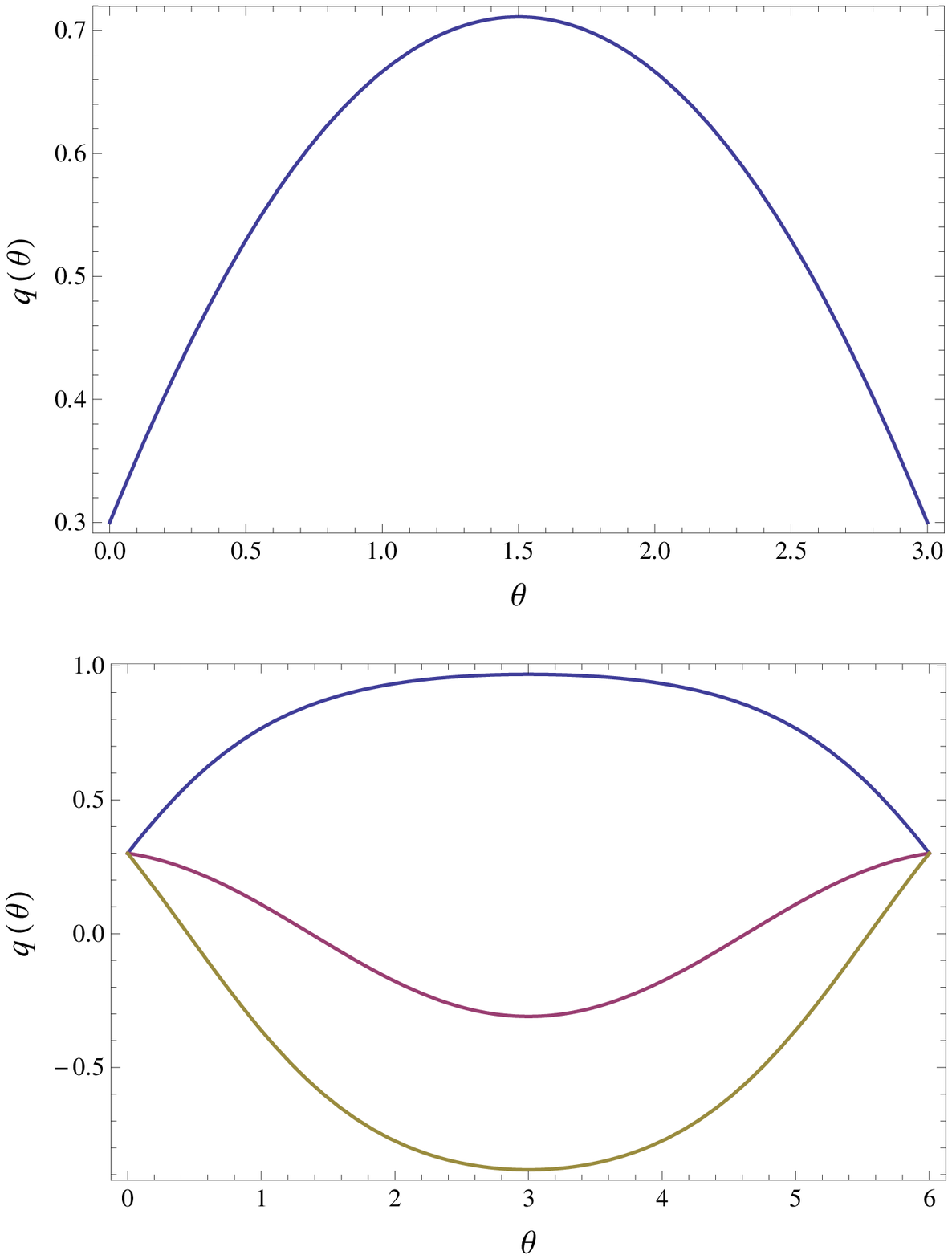}}
\end{center}
\vspace{-5mm}
\caption{\label{figsol}Plots of the classical solutions $q(\theta)$  for $q_0 = 0.3$ and $\Theta = 3$ (upper), before the first caustic, and for $\Theta = 6$ (lower), after the 
first caustic. In the latter, the middle curve corresponds to a (local) maximum of the action and, thus, will not be taken into consideration in the subsequent calculation of 
the partition function.}
\end{figure}

For those trajectories, the fluctuation determinant $\Delta$ defined in \eqref{determinant} can be expressed as 
\cite{deCarvalho:1998mv,deCarvalho:1998ff}:
\begin{equation}
\label{determinant2}
\Delta = \frac{4\pi g[U(q_t) - U(q_0)]}{U'(q_t)}\left(\frac{\partial\Theta}{\partial q_t}\right)_{q_0}
\end{equation}

\noindent
That being so, we can now use the previous equation in \eqref{semiclassZ} to obtain the semiclassical partition 
function.

The standard semiclassical method yields a very good approximation for the density matrix before and after the 
first caustic. However,  as discussed in Ref. \cite{deCarvalho:2001vk} (see also below), the method breaks down at the 
caustic, since, there, by construction, the determinant $\Delta$ vanishes --- see equation \eqref{semiclassZ}. This can be easily understood in the functional space (see Figs. \ref{figaction1} and \ref{figaction2}): the second derivative of the action vanishes whenever two (or three) solutions coalesce. Therefore, any approximation that 
stops at the quadratic term is bound to diverge at this point.

This singularity, however, is integrable (as also noted in \cite{deCarvalho:2001vk}). This statement can be proved if we perform a change of variables in
\eqref{semiclassZ} from $q_0$ to $q_t$. Using \eqref{ToF} and \eqref{determinant2} we can write, following
Ref. \cite{deCarvalho:1998mv}:
\begin{equation}
\left(\frac{\partial q_0}{\partial q_t}\right)_{\Theta} =  -\frac{U'(q_t)\Delta}{4\pi g v(q_0,q_t)} \ .
\end{equation}

\noindent
Thus, the standard semiclassical partition function is written as
\begin{equation}
Z = -\frac{1}{4\pi g}\sum_i \int_{q_{\Theta}^-}^{q_{\Theta}^+} dq_t \ \frac{U'(q_t)\Delta^{1/2}}{v(q_0,q_t)}\exp(-I[\bar{q}^{~i}_c]/g) ,
\end{equation}

\noindent
where $v(q_0,qt) \equiv \textrm{sign}(q_t-q_0)\sqrt{2[U(q_0)-U(q_t)]}$ and 
$q_{\Theta}^\pm \equiv \lim_{q_0 \rightarrow \pm \infty} q_t(q_0,\Theta)$. Therefore, the change of variables
removes the singularity and this procedure, summing over the two minima of the euclidean action, should give a reasonable
approximation to the partition function.

However, thermodynamic quantities are obtained taking derivatives of the partition function and thus they are affected by the
singularity. Therefore, as we are interested in computing the specific heat, we shall take our calculation up to the fourth order 
in the fluctuations. 
Notice that this is still a semiclassical expansion, for we assume that the main contribution comes from the classical solution. 
The calculation is depicted in Appendix \ref{appnormal}, where one 
can also promptly recognise the standard semiclassical expansion\footnote{It is also obvious then when this approximation fails: by neglecting terms of order $c_0^3$ and higher, Eq. (\ref{action_n}) yields the usual term $\Delta^{-1/2}$, which diverges  at the caustic. } if one stops at the second term on the right-hand side of Eq.~(\ref{action_n}). Nevertheless, even the full expression is not useful for practical purposes, for its calculation requires the knowledge of the eigenfunction $y_0(\theta)$ and its eigenvalue $c_0$. There is, however, a shortcut \cite{deCarvalho:2001vk}: just as in a plain 4th-order polynomial of the form (see Eq. (\ref{normalform}) and Table \ref{tablenormalform})
\begin{equation}
f(x) = \frac{1}{4} x^4 + \frac{a}{2} x^2 + b x + c\, ,
\end{equation}
the coefficients $\{a,b,c\}$ are completely determined by the values of the function $f(x)$ in 3 points. In the present case, all we need are the values of the action at the 3 extrema, easily calculated from the classical trajectories (\ref{jacobicd}).

\begin{figure}[ht]
\begin{center}
\resizebox*{!}{5.5cm}{\includegraphics{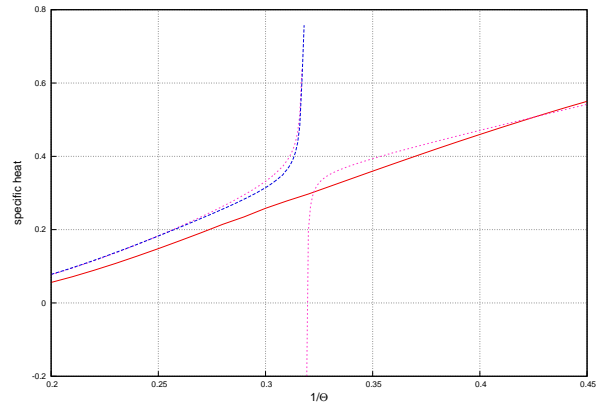}}
\end{center}
\vspace{-5mm}
\caption{\label{figspec_cat}Specific heat {\it vs} $1/\Theta$ for $g = 0.1$. Dotted/magenta line: only the global minimum is taken into account.  Dashed/blue line: both minima are taken into account.  Solid/red line: current approach.}
\end{figure}

\begin{figure}[ht]
\begin{center}
\resizebox*{!}{5.5cm}{\includegraphics{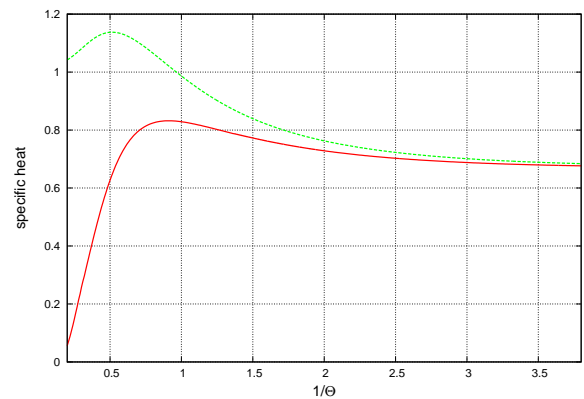}}
\end{center}
\vspace{-5mm}
\caption{\label{figspec_class}Specific heat {\it vs} $1/\Theta$ for $g = 0.1$. Green/dashed: classical result; Red/solid: current approach.}
\end{figure}

The following plots present the results obtained for the specific heat for $g=0.1$ --- we argue in the next subsection that this value is small enough so that the  effects of the catastrophe are relevant.  In Fig. \ref{figspec_cat} we show the results for the specific heat from different approaches around the temperature where the first catastrophe takes place (at $\Theta = \pi$). On the right-hand side (higher-temperature, lower-$\Theta$) of the plot, the action has only 1 minimum; on the left-hand side (lower-temperature, higher-$\Theta$), the action has 2 minima. Accordingly, we plot (dotted/magenta line) the standard semiclassical approximation around the global minimum of the action, (dashed/blue line) the standard semiclassical approximation around both minima of the action, considered independent and far apart from each other, and (solid/red line) the current approach. The former two calculations are supposed to diverge at the catastrophe due to the coalescence of the classical trajectories. Our results are also to compared the classical one in Fig.~\ref{figspec_class}.

Here, we show the corrections that should be taken into account when the standard semiclassical approximation fails, going beyond second order in the perturbation whenever this approximation yielded divergent results. On the other hand, we still rely on the assumption that the classical solution is responsible for the main contribution to the partition function (and, consequently, to relevant thermodynamic quantities, such as the specific heat). In other words, we assume throughout the paper that the first term of the JWKB expansion of the partition function is a good approximation. 

Just as in any standard quantum mechanics calculation, one does not expect the JWKB approximation to hold when the thermal energy is close to the height of the barrier, where the potential changes quickly and the classical turning points are too close to each other. 
 Therefore, one must require here that $E_b/E_T \equiv  \Theta/(4g) \gg 1$, where $E_b\equiv V(0) - V(\pm \sqrt{mw^2/\lambda})$ is the height of the barrier and $E_T\equiv 1/\beta \equiv k_B T$ corresponds to the thermal energy. In other words, unless $g\ll \Theta_c/4\sim 0.8 $, the JWKB approximation itself will break down before the first catastrophe sets in at $\Theta_c=\pi$.

\section{Summary}

Semiclassical approximations usually uncover important non-perturbative information about quantum systems. They are specially 
suited to the construction of effective theories at finite temperature, since the perturbative approach suffers from serious infrared 
problems and needs involved resummation techniques to provide sensible results. The boundary effective theory has proved to 
be very adequate to describe the thermodynamics of a thermal massless scalar theory, providing an excellent result for the pressure 
at leading order \cite{Bessa:2010tj}, as well as a consistent description of the thermal effective potential in the symmetric 
sector \cite{Bessa:2011tn}. Its extension to the case where spontaneous symmetry breaking is present is, nevertheless, 
subtle, the main obstacle being the existence of multiple extrema of the action for sufficiently low temperatures and the associated 
bifurcations of classical solutions.

In this paper we have considered, as a toy model (however not an academic one, since the double well has, of course, applications 
in statistical mechanics and condensed matter physics), the analogous case in quantum statistical mechanics. We have shown how 
to use the tools of catastrophe theory to deal with caustics and provide finite and well-behaved results for the partition function and 
the specific heat. In particular, we have proved that one needs at most two relevant classical solutions in the procedure, which renders 
the method of practical use. As mentioned previously, this corresponds to a first step towards the study of spontaneous symmetry 
breaking and thermal phase transitions using the boundary effective theory, on which we plan to report soon.

%%%%%%%%%%%%%%%%%%%%%%%%%%%%%%%%%%%%%%%

\section*{Acknowledgments}

This work was partially supported by CAPES-COFECUB project 663/10, CNPq, FAPERJ, FAPERN, FUJB/UFRJ and ICTP. 
The work of ESF was financially supported by the Helmholtz International Center for FAIR within the framework of the LOEWE 
program (Landesoffensive zur Entwicklung Wissenschaftlich-\"Okonomischer Exzellenz) launched by the State of Hesse.

%%%%%%%%%%%%%%%%%%%%%%%%%%%%%%%%%%%%%%%

\appendix
\section{Elements of Catastrophe Theory}\label{appcatastrophe}

Consider a function $S(x;\nu)$ that depends on a set of coordinates $x = \{x_1,x_2,...\}$ and certain control
parameters $\nu = \{\nu_1,\nu_2,...\}$. The number of coordinates is the dimension of the catastrophe, while the number of control parameters defines the so-called codimension of the catastrophe.

In two dimensions, $S$ can be seen, for instance, as describing the 
terrain height of a certain landscape. Its maxima, minima and saddle points represent the peaks, valleys and throats. 
In this picture, the role of the parameters $\nu$ is to deform the topography of the landscape, 
changing the position of the extrema and eventually splitting or merging some of them.

The aim of catastrophe theory is to study how the pattern of the so-called generating function $S$ is qualitatively altered when the 
control parameters are changed. Within this framework, one is able to understand how the extrema coalesce and separate as the parameters $\nu_k$ are varied, in a systematic and quite general approach.

Catastrophe theory \cite{catastrophe} characterizes the stable singularities
under changes in the generating functional $S$: those are the so-called elementary catastrophes.
The splitting lemma \cite{catastrophe} guarantees that it is always possible to write such stable generating functions in their normal forms, according to Table \ref{tablenormalform}. They can also be arranged hierarchically
%, as shown in table \ref{hierac}
: whenever a given catastrophe is identified, all of its subordinated ones --- those with the same dimension and smaller codimension --- will also be present. 

Let us consider the phase space $(x,\nu)$ defined by both the coordinates and control parameters of the function $S$. Obviously, the locus of the extrema, the so-called equilibrium surface, of $S$ is given by
\begin{equation}
\frac{\partial S}{\partial x_i}(x_e,\nu) = 0 \ .
\label{eqsurf}
\end{equation}
\noindent
i.e., if for certain values of the control parameters $\nu$, the point $x_e$ represents an extremum of $S$, and the point $(x_e,\nu)$ is said to lie on the equilibrium surface. 

% \begin{table}
% \begin{tabular}{lcr}
% swallowtail & hyperbolic umbilic & elliptic umbilic \\
% ~ & cusp & ~\\
%  ~& fold & ~
% \end{tabular}
% \caption{Hierarchical arrangement of the first elementary catastrophes. Whenever a given catastrophe is identified, the ones on the lines below are also present.}
% \end{table}

Note, however, that no information about the nature of the extrema is given by 
\eqref{eqsurf}. In order to determine whether a given extremum is a minimum, maximum or a saddle-point, one has to study the eigenvalues of the Hessian matrix $H$ calculated at the equilibrium points $x_e$, whose elements 
are defined as
\begin{equation}
H_{ij} = \left. \frac{\partial^2 S}{\partial x_i \partial x_j}\right|_{x_e} \ .
\label{hessian}
\end{equation}
\noindent
When all the eigenvalues of $H$ are positive, we have a minimum; when all are negative, a maximum and when $k$ are negative and the others positive, the extremum under consideration is a $k$-saddle-point. 

It is clear from the previous considerations that the nature of the extrema changes when some of the related 
eigenvalues of  $H$ change sign. Therefore, if none of the eigenvalues is zero, a small change of the parameters will not affect the nature of the extrema.

At the bifurcation set, when one or more of the eigenvalues vanish, the situation changes drastically, as any small change of the control parameters will make the eigenvalue(s) positive or negative, changing the nature of the extremum. 
In other words, the qualitative aspect of the function is changed whenever the 
determinant of $H$ vanishes.

One can see such behavior clearly present in Fig. \ref{figcuspid}, which represents the bifurcation set in the control parameter space, with codimension 2: crossing at the vertex correponds to the coalescence of 3 extrema (Fig. \ref{figaction1}): this is the cusp catastrophe. Along the bifurcation set, however, there is only one free control parameter (codimension 1) --- since Eq. (\ref{bifeq}) introduces a constraint between the two of them. On this curve, only 2 trajectories coalesce (Fig. \ref{figaction2}): this is the fold, subordinated to the cusp. 

In the next Appendix, we show how one can write the action in the normal form corresponding to the cusp. 

\begin{table}
\begin{tabular}{cccc}
catastrophe & codim & dim & normal form\\
\hline
fold & 1 & 1 &  {\tiny ${x^3}/{3} + u x$}	\\
cusp & 2 & 1 &  {\tiny $x^4/4 + u {x^2}/{2} + v x$}	\\
swallowtail & 3 & 1 & {\tiny ${x^5}/{5} + u {x^3}/{3} + v {x^2}/{2} + w x$}	\\
elliptic umbilic & 3 & 2 &  {\tiny $x^3 -3xy^2 - u (x^2 + y^2) - v x - w x$}	\\
hyperbolic umbilic & 3 & 2 & {\tiny $x^3 + y^3 + u x y - vx - wy$}
\end{tabular}
\caption{The five simplest elementary catastrophes, their codimensions (number of control parameters), dimensions (number of coordinates) and the normal forms of their generating functions.}
\label{tablenormalform}
\end{table}

%%%%%%%%%%%%%%%%
\section{The normal form of the action}\label{appnormal}
In this section, we show how the action can be written in normal form, as in Eq.~(\ref{normalform}). 

In the first place, we write $q(\theta) = q_{cl}(\theta) + \eta(\theta)$, so that the euclidean action is cast in 
the form:
\begin{align}
I[q(\theta) + \eta(\theta)] &= I[q_{cl}(\theta)] \nonumber \\
&+\frac{1}{2}\int^{\Theta}_0\eta(\theta)\left[-\frac{d^2}{d\theta^2} 
-1+3q^2_{cl}(\theta)\right]\eta(\theta)d\theta\nonumber \\
&+ \int^{\Theta}_0\left[q_{cl}(\theta)\eta^3(\theta) + \frac{1}{4}\eta^4(\theta)\right]d\theta \ .
\label{action_q_eta}
\end{align}

Notice that the classical solution was not specified. There are two interesting cases: the identically null function 
($q_{cl} \equiv 0$), or one of the new functions. In the latter case, the calculation is obviously made after they
appear.

Now, we expand the perturbation $\eta(\theta)$ in terms of the eigenfunctions of the fluctuation operator,
i.e. in terms of the functions $y_j(\theta)$ satisfying the following equation:
\begin{equation}
\left[-\frac{d^2}{d\theta^2} -1+3q^2_{cl}(\theta)\right]y_j(\theta) = \alpha_jy_j(\theta) \ .
\end{equation}

\noindent
The eigenfunctions can be taken as orthonormal in the interval $[0,\Theta]$:
\begin{equation}
\int^{\Theta}_0y_i(\theta)y_j(\theta)d\theta = \delta_{ij} \ .
\end{equation}

\noindent
Furthermore, they must satisfy the following boundary conditions
\begin{equation}
y_j(0) = y_j(\Theta) = 0 \ \forall j \ .
\end{equation}

\noindent
Expanding $\eta(\theta)$ in terms of $y_j(\theta)$, we have
\begin{equation}
\eta(\theta) = \sum^{\infty}_{j=0}c_jy_j(\theta) \ .
\label{eigen}
\end{equation}

Thus, using the expansion of the fluctuations and the orthonormalization conditions, we can write the action as:
\begin{align}
I =& I_{cl} + \frac{1}{2}\sum_jc^2_j\alpha_j+\sum_{ijk}c_ic_jc_k\int^{\Theta}_0q_{cl}y_iy_jy_k \ d\theta \nonumber \\ 
&+\frac{1}{4}\sum_ic^4_i \ .
\end{align}

We have to impose the fact that the classical solution $q_{cl}$ is an extremum of the action, therefore the 
fluctuations vanish, i.e. $c_j = 0$, at $q_{cl}$ . Equivalently:
\begin{equation}
\left.\frac{\partial I}{\partial c_i}\right\vert_{c_i=0} = 0 \ .
\end{equation}

\noindent
This leads to the following expression for the action:
\begin{equation}
I \approx I_{cl} + \frac{\alpha_0}{2}c^2_0 + c^3_0\int^{\Theta}_0 q_{cl}y^3_0 \ d\theta + \frac{1}{4}c^4_0 + 
\sum_{j \neq 0} \frac{\alpha_j}{2}c^2_j \ .
\label{action_n}
\end{equation}

\noindent
In the previous equation, $j = 0$ denotes the eigenfunction whose eigenvalue is about to vanish. Besides, we have 
neglected terms of the order $c^3_j$ for $j \neq 0$.

The difference between this expression and the usual saddle-point approximation is the inclusion of higher-order terms
in the variable $c_0$, the one related with the vanishing eigenvalue, while only terms up to second order in the 
other variables, related with the other directions in functional space.

Now we perform the following change of variables:
\begin{subequations}\label{eulerlagrange}
\begin{gather}
z \equiv c_0 + \Upsilon\\
u \equiv \alpha_0 -3\Upsilon^2\\
v \equiv \Upsilon(2\Upsilon^2 - \alpha_0)\\
s \equiv I_{cl} + \frac{\Upsilon^ 2}{2}\left(\alpha_0 - \frac{3}{2}\Upsilon^2\right)\\
\Upsilon \equiv \int^{\Theta}_0 q_{cl}y^3_0 \ d\theta
\end{gather}
\end{subequations}

\noindent
allowing us to write the action in the so called normal form, as in Eq. \eqref{normalform}.

%%%%%%%%%%%%%%%%%%%%%%%%%%%%%%%%%%%%%%%

\end{document}